\documentclass[12pt]{article}

\usepackage{a4wide}
\usepackage{graphicx}
\usepackage{epsfig}
\usepackage{amsmath}
\usepackage{amsfonts}
\usepackage{amssymb}
\usepackage{dsfont}
\usepackage{pstricks}
\usepackage{pstricks-add}
\usepackage{enumerate}
\usepackage{psfrag,colordvi}

\usepackage{lineno}
\usepackage{subfigure}

\newtheorem{proposition}{Proposition}[section]

\newtheorem{theorem}{Theorem}[section]
\newtheorem{remark}{Remark}[section]

\def\endpf{\hfill$\blacksquare$}

\newcommand{\ls}[1]
   {\dimen0=\fontdimen6\the\font \lineskip=#1\dimen0
\advance\lineskip.5\fontdimen5\the\font \advance\lineskip-\dimen0
\lineskiplimit=.9\lineskip \baselineskip=\lineskip
\advance\baselineskip\dimen0 \normallineskip\lineskip
\normallineskiplimit\lineskiplimit \normalbaselineskip\baselineskip
\ignorespaces}
\ls{0.95}

\newcommand{\T}{\mathbf{T}}
\newcommand{\x}{\mathbf{x}}
\newcommand{\scal}[1]{\langle #1 \rangle}

\usepackage{sidecap}

\begin{document}

\title{Continuum Equilibria and Global Optimization \\ for Routing in
Dense Static Ad Hoc Networks}
\author{
Alonso Silva\footnote{INRIA,
B.P.93, 2004 Route des Lucioles, 06902 Sophia-Antipolis Cedex,
France. Email: \texttt{\{alonso.silva, eitan.altman\}@sophia.inria.fr}}~$^{\ddag}$~,
Eitan Altman$^*$,
Pierre Bernhard\footnote{I3S, Universit\'e de Nice-Sophia Antipolis
and CNRS, 940 Route des Colles, B.P. 145, 06903 Sophia-Antipolis
Cedex, France. Email: \texttt{pierre.bernhard@polytech.unice.fr}}~,
M\'erouane Debbah\footnote{
Alcatel-Lucent Chair in Flexible Radio - SUPELEC, 91192 Gif sur Yvette, France.
Email: \texttt{merouane.debbah@supelec.fr}}
}
\date{}
\maketitle

{\small
{\bf Abstract}
We consider massively dense ad hoc networks
and study their continuum limits as the node density increases and as
the graph providing the available
routes becomes a continuous area with location and congestion
dependent costs. We study both the global optimal solution as
well as the non-cooperative routing
problem among a large population of users where
each user seeks a path from its origin to its destination
so as to minimize its individual cost. Finally, we seek for a
(continuum version of the) Wardrop equilibrium.  We first
show how to derive meaningful cost models as a function of the
scaling properties of the capacity of the network and of
the density of nodes.
We present various solution methodologies for the problem: (1) the
viscosity solution of the Hamilton-Jacobi-Bellman equation,
for the global optimization problem,
(2) a method based on Green's Theorem for the least cost problem of an
individual, and (3) a solution of the Wardrop equilibrium problem
using a transformation into an equivalent global optimization problem.
}

\bigskip


\ls{0.95}

\section{Introduction}

\begin{figure}[tbh]
\centering
\includegraphics[angle=90,height=5in,width=6.5in ]{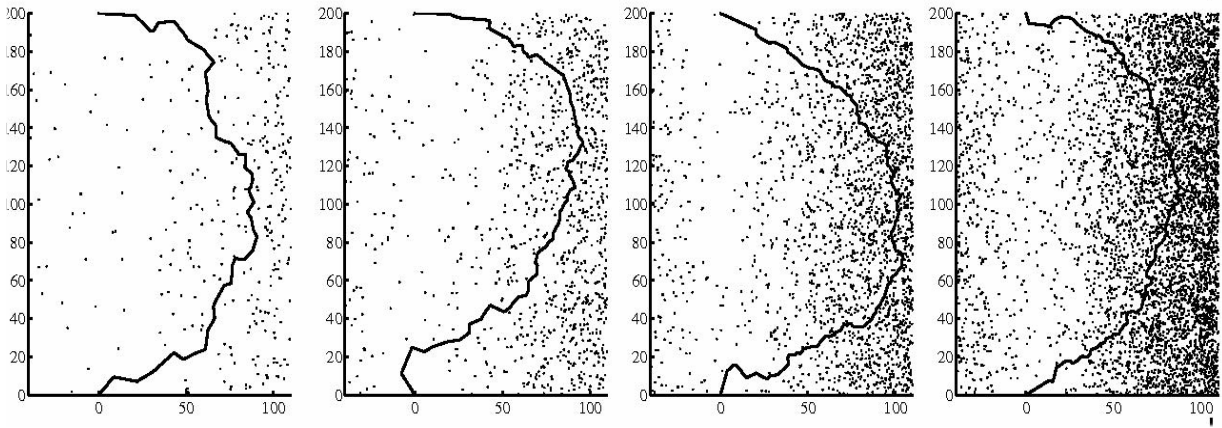}
\vspace{-3cm}

 \caption{Minimum cost routes in increasingly large networks.}
\label{figOpt}
\end{figure}

Research on ad hoc networks involves the design of protocols 
at various network layers
(MAC, transport, etc.),
the investigation of physical limits of transfer rates,
the optimal design of end-to-end routing, efficient energy management,
connectivity and coverage 
issues, 
performance analysis of delays,
loss rates, etc. The study of these issues has required the use of
both engineering methodologies as well as information theoretical ones,
control theoretical tools, queueing theory, and others. 
One of the most challenging problems in the performance analysis and in the
control of ad hoc networks has been routing in massively dense ad hoc networks.
On one hand, when applying existing tools for optimal routing, the complexity 
makes the solution intractable as the number of nodes becomes very large.
On the other hand, it has been observed that as an ad hoc network becomes
``more dense" (in a sense that will be defined precisely later), the optimal
routes seem to converge to some limit curves. This is illustrated in
Fig.~\ref{figOpt}. We call this 
regime,
the limiting ``macroscopic'' regime.
We shall show that the solution to the macroscopic behavior 
({\it i.e.}, the limit of the optimal routes as the system becomes more and
more dense) 
is sometimes 
much easier to solve than the original
``microscopic model''.

\vspace{-3cm}

The term ``massively dense'' ad hoc networks is used to indicate
not only that the number of nodes is large, but also
that the network is highly connected.
By the term ``dense'' we further understand that for every point
in the plane there is a node close to it with high probability; by
``close" we mean that its distance is much smaller than the transmission
range. In this paper and in previous works (cited in
the next paragraphs) one actually studies the limiting properties
of massively dense ad hoc networks, as the density of nodes tends
to infinity.


The empirical discovery of the macroscopic limits motivated a large
number of researchers to investigate continuum-type limits of the
routing problem. A very basic problem in doing  so has been
to identify the most appropriate scientific context for modelling
and solving this continuum limit routing problem.
Our major
contribution is to identify completely the main paradigms
(from optimal control as well as from road traffic engineering)
for the 
modelling 
and the solution of this problem. We illustrate the use
of these methodologies by considering new types of models that arise
in the case of nodes with directional 
antennas.

{\bf Physics-inspired paradigms:}
The physics-inspired paradigms used for the study of large
ad hoc networks go way beyond those related to statistical-mechanics
in which macroscopic properties are derived from microscopic structure.
Starting from the pioneering work by Jacquet (see \cite{geometry})
in that area, a number of research groups have worked on
massively dense ad hoc networks using tools from
geometrical optics \cite{geometry}\footnote{We note that this
approach is restricted  to costs that do not depend on the
congestion.}. 
Popa {\it et al.} in \cite{popa1} studied optical paths and actually showed that 
the optimal solution to a minmax problem of load balancing 
can be achieved by using an appropriately chosen optical 
profile. The forwarding load appears to correspond to the 
scalar sum of traffic flows of different classes. This 
means that the optimal solution (with respect to this objective) 
can be achieved by single path routes, a result obtained 
also in \cite{popa2}. Similar problems have been also studied in 
%
\cite{popa4}, as well as in works doing load balancing by analogies to Electrostatics
(see {\it e.g.}~\cite{GT,KS1,KS2,TT,design}, and
the survey \cite{toumpis} and references therein). We shall describe
these in the next sections.

The physical paradigms allow the authors
to minimize various metrics related to the routing problem. In contrast,
Hyytia and Virtamo proposed in \cite{jorma} an approach based on load
balancing arguing that if shortest path (or cost minimization)
arguments were used, then some parts of the network would carry
more traffic than others and may use more energy than others. This
would result in a shorter lifetime of the network since some
parts would be out of energy earlier than others.


{\bf Road-traffic paradigms:}
The development of the original
theory of routing in massively dense networks among the
community of ad hoc networks has emerged in a complete independent
way of the existing theory of routing in massively dense networks
which had been developed within the community of road traffic engineers.
Indeed, this approach had already been introduced in 1952
by Wardrop \cite{Wardrop} and by Beckmann \cite{Beckmann} and
is still an active research area among that community, see
\cite{Dafermos,daniele02,hoWong,idone04,wongsens} and references therein.

{\bf Our contribution and the paper's structure:}
We combine in this paper various approaches from the area
of road traffic engineering
as well as from optimal control theory in order to formulate
models for routing in massively dense networks.
We further propose a simple novel approach
to that problem using a classical device of 2-D, singular optimal
control \cite{Miele}, based on Green's formula
to obtain a simple characterization of
least cost paths of individual packets. We end the paper by 
a numerical example for computing an equilibrium.

The paper  starts with a background on the research
on massively dense ad hoc networks. In 
doing so, it is not limited to a 
specific structure of the cost. However, when introducing our approach based
on road traffic tools, we choose to restrict ourselves to 
static networks (say sensor networks) having a special cost structure
characterized by communications through horizontally and vertically oriented
directional antennas. The use of directional antennas, by pointing information
in a specific direction, allows one
to save energy which may result in
a longer life time of the network. The nodes are assumed to be placed
deterministically. For an application of our approach to
omnidirectional antennas, see \cite{road1}. We solve various
types of optimization problems: We consider (i) the global optimization
problem in which the objective of routing decisions is to minimize
a global cost, (ii) the individual optimization problem
whose solution is the Wardrop equilibrium. It
corresponds to the situation where the number of users is very large,
and each user tries to minimize its own cost (in a non-cooperative way).
This is a ``population game'' or a ``non-atomic-game'' framework.
It is called ``non-atomic'' since a single player sends a 
negligible amount of traffic (with respect to the total 
amount of traffic) and as a consequence, its impact on the performance
of other users is negligible. 

The structure of this paper is as follows: We begin by
presenting models for costs relevant to
optimization models in routing or to node assignment.
We then formulate the global optimization problem
and the individual optimization one with a focus on the
directional antennas scenario. We provide several
approaches obtaining both a qualitative characterization
as well as quantitative solutions to the problems.

\section{An overview of dense ad hoc networks}

We suppose that the network is modeled in the two dimensional plane $X_1\times X_2$. 
The continuous {\bf information density function} $\rho(\mathbf x)$,
measured in $\textrm{bps}/\textrm{m}^2$, at locations $ \mathbf x $
where $\rho( \mathbf x )>0$ corresponds to a distributed data origin such that the
rate with which information is created
in an infinitesimal area of size $d A_\varepsilon $ centered at
$ \mathbf x $ is $\rho( \mathbf x )\,d A_\varepsilon $. Similarly, at locations
$ \mathbf x $ where $\rho( \mathbf x )<0$
there is a distributed data sink such that the rate with which
information is absorbed by an infinitesimal area of size $d A_\varepsilon$, centered at point $ \mathbf x $,
is equal to $-\rho(\mathbf x )\,d A_\varepsilon$.

The total rate at which the destination nodes receive the data must
be the same as the total rate at which the data
is created at the origin nodes, {\it i.e.},
\begin{equation*}
\int\limits_{X_1\times X_2}\rho( \mathbf x )\,dS=0.
\end{equation*}

In optimizing a routing protocol in ad hoc networks, or in
optimizing the placement of nodes, one of the starting
points is the determination of the cost function that captures the cost of transporting a packet through the network.
To determine it, we need a detailed specification of the network which includes the following:
\begin{itemize}
\item A model for the placement of nodes in the network.
\item A forwarding rule that nodes will use to select the next hop of a packet.
\item A model for the cost incurred in one hop, {\it i.e.}, for transmitting a packet to an intermediate node.
\end{itemize}
Below we present several ways of choosing cost functions.

\begin{SCfigure}\label{arrowflow}
\psfrag{T}{$\mathbf T(\mathbf x)$}
\psfrag{T1}{$T_1(\mathbf x)$}
\psfrag{T2}{$T_2(\mathbf x)$}
\psfrag{X1}{$X_1$}
\psfrag{X2}{$X_2$}
\psfrag{e}{$d\ell$}
\centering\includegraphics[height=2.0in,width=2.2in ]{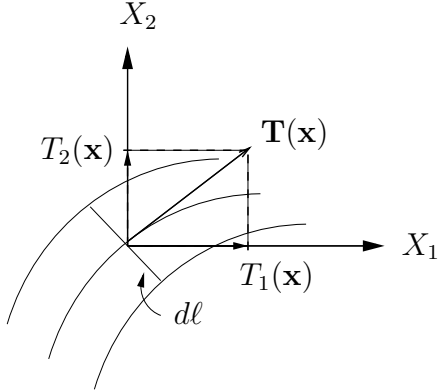}
\hspace{0.5 cm}

\caption{Flow of information~$\mathbf T(\mathbf x)$ through incremental line segment~$d\ell$,
decomposed in its horizontal component $T_1(\mathbf x)$~(in the direction~$X_1$) and its
vertical component $T_2(\mathbf x)$~(in the direction~$X_2$).}
\end{SCfigure}

We define the flow of information~$\mathbf T(\mathbf x)$ (see Fig.~\ref{arrowflow})
to be a vector whose components are the horizontal and vertical
flows at point $\mathbf x$. Throughout we assume that each point carries
a single flow (although the methodology can be extended to the multiflow case).
The restriction to a single flow is justified when there is either a single
destination, or when there is a set of destination points 
and the routing protocol has the freedom to decide to which of the set
the packets will be routed. Under this type of conditions, one may
assume a single flow at each point without loss of optimality
(see {\it e.g.}~\cite{popa2}).

\subsection{Costs derived from capacity scaling}
\label{costCap}

Many models have been proposed in the literature that show how
the transport capacity scales with the number of nodes~$n$
or with the density of nodes~$\lambda$ within a certain region.
A~typical cost (see {\it e.g.}~\cite{TT}) considered at a neighborhood of a location\footnote{We denote the vectors by bold fonts.}~$\mathbf x$
is the density of nodes required there to carry a given 
flow of information~$\mathbf T(\mathbf x)$.
We will work within a general framework and then investigate 
deeply some particular cases
with different possible protocols.
Assume that we use a protocol that provides a transport capacity of
the order of~$f(\lambda)$ at some region in which the density of nodes
is~$\lambda$ (we will provide examples of the function~$f$ ahead).
This means that in order to support a flow of information~$\lvert\mathbf T(\mathbf x)\rvert$
passing through a neighborhood of the location~$\mathbf x$,
we will need to place deterministically the nodes according to
the formula~$f^{-1}(\lvert\mathbf T(\mathbf x)\rvert)$.
Then if we assume that a flow of information $\mathbf T(\mathbf x)$~is assigned through a neighborhood of the location
the cost will be taken as
\begin{equation}\label{costEq}
c(\mathbf x, \mathbf T(\mathbf x))=f^{-1}(\lvert\mathbf T(\mathbf x)\rvert)
\end{equation}
where $\lvert\cdot\rvert$ represents the norm of a vector.

Most work in this area has considered the $\ell^2$-norm, {\it i.e.}, for
$\mathbf x = ( x_1 , x_2 )$, we define
\mbox{$\lvert\mathbf x\rvert = \sqrt{ x_1^2 + x_2^2 }$}.
We shall consider later also the $\ell^1$-norm, {\it i.e.}, $\lvert\mathbf x\rvert = \lvert x_1\rvert + \lvert x_2 \rvert$.

Examples for $f$:
\begin{itemize}
\item
Using a network theoretic approach based on multi-hop communication,
Gupta and Kumar proved in~\cite{gupta99}
that the throughput of the system that can be transported by the
network when the nodes are optimally located is\footnote{
We denote $f\in\Omega(g)$ if $f$ is bounded below by $g$ (up to a constant factor) asymptotically
and we denote $f\in \Theta(g)$ if $f$ is bounded both above and below by $g$ (up to a constant factor) asymptotically.} $\Omega(\sqrt\lambda)$,
and when the nodes are randomly located this throughput becomes
$\Omega(\frac{\sqrt \lambda }{\sqrt{\log \lambda}})$.
Using percolation theory, the authors of \cite{closing gap}
have shown that in the randomly located set the same $\Omega(\sqrt\lambda)$
can be achieved.
\item
Baccelli, Blaszczyszyn and M\"uhlethaler introduce in \cite{BBM}
an access scheme, MSR (Multi-hop Spatial Reuse Aloha),
reaching the Gupta and Kumar bound
$\Theta(\sqrt{\lambda})$ which does not require prior knowledge
of the node density.
\end{itemize}
We conclude that for the model of Gupta and Kumar
with either the optimal location or the random location approaches,
as well as for the MSR protocol with a Poisson distribution of nodes,
we obtain a quadratic cost of the form
\begin{equation}
\label{te}
c(\mathbf T(\mathbf x))=k \lvert\mathbf T(\mathbf x)\rvert^2
=k ( T_1( {\bf x } )^2 + T_2( {\bf x } )^2 ).
\end{equation}
This follows from the fact that in the previous examples
$f(x)$ behaves like $\sqrt{x}$, so the inverse of the function~$f$
must be quadratic. Then from \eqref{costEq} we conclude that
the cost function must be quadratic on~$\lvert\mathbf T(x)\rvert$.


\subsection{Congestion independent routing}

A metric often used in the Internet for determining routing costs
is the number of hops from origins to destinations, which routing protocols try to
minimize. The number of hops is proportional to the expected
delay along the path in the context of ad hoc networks,
in case the queueing delay is negligible with respect
to the transmission delay over each hop.  This criterion is insensitive
to interference or congestion. We  assume that it depends only on
the transmission range.
We describe various cost criteria that can be formulated with this approach.
\begin{itemize}
\item
If the range is constant then the cost density $c(\mathbf x)$ is constant so
that the cost of a path is its length in meters. The
routing then follows a shortest path selection.
\item
Let us assume that the range $R(\lambda;\mathbf x)$ is small, and it
depends on local radio conditions
at position $ \mathbf x $ (for example, if it
is influenced by weather conditions) but not
on interference. The latter is justified when
dedicated orthogonal channels ({\it e.g.}~in time or frequency) can be allocated
to traffic flows that would otherwise interfere with each other.
Then determining the optimal routing becomes a path cost minimization problem.
We further assume, as in~\cite{gupta99}, that the range is scaled
to go to~$0$ as the total density $\lambda$
of nodes grows to infinity. More precisely,
let us consider a scaling of the range such that the
following limit exists:
\begin{equation*}
r(\mathbf x):=\lim_{\lambda\to\infty}\frac{ R(\lambda;\mathbf x) }{  \lambda }
\end{equation*}
Then in the dense limit, the fraction of nodes that participate
in  forwarding packets along a path is $1/r( \mathbf x )$ at position~$\mathbf x$,
and the path cost is the integral of this density along the path.
\item
The influence of varying radio conditions on the  range can be
eliminated using power control that can equalize the hop distance.
\end{itemize}

\subsection{Costs related to energy consumption}

In the absence of capacity constraints, the cost
can represent energy consumption.
In a general multi-hop ad hoc network, the hop distance can be optimized
so as to minimize the energy consumption.
Even within a single
cell of 802.11 IEEE wireless LAN one can improve the
energy consumption by using multiple hops,
as it has been shown not to be efficient in terms of
energy consumption to use a single hop \cite{venkatesh}.

Alternatively, the cost can take into account the scaling
of the nodes (as we have done in~Section \ref{costCap}) that is
obtained when there are energy constraints. As an example,
assuming random deployment of nodes, where each node has
data to send to another randomly selected node, the capacity
(in bits per Joule) has the form
$f(\lambda) = \Omega\left((\lambda/\log \lambda)^{(q-1)/2} \right) $
where $q$ is the path-loss, see \cite{rodoplu}.
The cost is then obtained using (\ref{costEq}).

\section{Preliminaries}
\label{sec:elect}
\index{Wireless Sensor Networks!Optimal Deployment}

In the work of Toumpis {\it et al.}~(\cite{GT,TT05,TT, design, toumpis, CMT}),
the authors addressed the problem
of the optimal deployment of wireless sensor networks by 
a parallel with Electrostatics. We shall recall below 
the representation of the flow conservation constraint,
which is well known in Electrostatics. This derivation
appears both in physics-inspired papers~(\cite{GT,TT05,TT, design, toumpis, CMT}) as well as in
the road traffic literature \cite{Dafermos}.

Consider a grid area network
\renewcommand{\Omega}{D}
$\Omega_0$ of arbitrary shape on the two-dimensional plane with axis $X_1$ and axis $X_2$,
with smooth boundary.
It is necessary that the rate with which information is created in the area must be equal to the
rate with which information is leaving that area, {\it i.e.},
\begin{equation}\label{integralgauss}
\int\limits_{\Omega_0}\rho( \mathbf x )\,d\mathbf x=
\oint\limits_{\partial \Omega_0}[\mathbf{T}(\x)\cdot\mathbf{n}(\x)]\,d\ell.
\end{equation}

The integral on the left is the surface integral of $\rho(\mathbf x )$ over $\Omega_0$. The integral on the right is the path integral of
the inner product $\mathbf T\cdot\mathbf n$ over the curve $\partial\Omega_0$. The vector $\mathbf n(\x)$ is the unit normal vector 
to $\partial \Omega_0$ at the boundary point $\x\in\partial \Omega_0$ and pointing outwards. The function $\mathbf T(\x)\cdot\mathbf n(\x)$,
measured in $\textrm{bps}/\textrm{m}$, is equal to the rate with which information is leaving the domain $\Omega_0$ per unit length of boundary at the boundary point $\x$.

As this holds for \emph{any} (smooth) domain $\Omega_0$, it follows that necessarily 
\begin{equation}\label{div}
\nabla\cdot\mathbf T(\x):=
\frac{\partial {T}_1 (\mathbf x) }{ \partial x_1 } +
\frac{\partial {T}_2 (\mathbf x) }{ \partial x_2 } =
\rho(\x),
\end{equation}
where ``$\nabla\cdot$'' is the divergence operator.
Notice that equations~\eqref{integralgauss} and~\eqref{div} are the integral
and differential versions of Gauss's law, respectively.

{\bf Extension to multi-class traffic}
The work on massively dense ad hoc networks
considers a single class of traffic.
In the geometrical optics approach it corresponds
to demand from a location~$\mathbf a$ to a location~$\mathbf b$. In the 
Electrostatic case
it corresponds to a set of origins and a set of destinations where
traffic from any origin point could go to any destination point.
The analogy to positive and negative charges in Electrostatics
may limit the perspectives of multi-class problems where traffic
from distinct origin sets has to be routed to distinct destination sets.

The model based on geometrical optics can directly be extended to
include multiple classes as there are no elements in the model that
suggest coupling between classes. This is due in particular to the
fact that the cost density has been assumed to depend only on the
density of the nodes and not on the density of the flows.

In contrast, the cost in the model based on Electrostatics
is assumed to depend both on the location as well as on the
local flow density. It thus models more complex interactions
that would occur if we considered the case of $\nu$ traffic
classes.  Extending the relation (\ref{div}) to the
multi-class case, we have traffic conservation at each point in space
for each traffic class as expressed in the following:
\begin{equation}\label{conservation}
\mathbf{\nabla}\cdot\mathbf T^j (\mathbf x) =
\rho^j (\mathbf x), \qquad\forall\mathbf x\in \Omega.
\end{equation}
The function $\mathbf T^j$ is the flow distribution of class $j$ and $\rho^j$ corresponds to
the distribution of the external origin and/or destinations.

Let $\mathbf T(\mathbf x)$ be the
total flow vector at point $ \mathbf x \in \Omega$.
It is a vector of dimension $\nu$, and each one of the
$ \nu$-entries is a two dimensionnal flow.
A generic multi-class optimization problem would then be:
minimize $Z$ over the flow distributions $\{\mathbf{T}^j\}$
\begin{equation}
Z=\int_\Omega c(\mathbf x,\mathbf T(\mathbf x))\,d\mathbf{x} 
\quad\mbox{subject to}\quad
\mathbf{\nabla}\cdot\mathbf T^j (\mathbf x) =\rho^j (\mathbf x)
, \ j=1,...,\nu
\qquad \forall \mathbf x \in \Omega.
\label{genOpt}
\end{equation}

\section{Directional Antennas and Global Optimization}
\label{direct}

So far we have adopted a general framework 
under which the flow is conserved. To proceed, we need more specific
assumptions on the cost function. The one we shall introduce
here can be called an $\ell^1$-norm model, in which the cost to go from
a point to another is the sum of the horizontal and vertical cost
components. This is justified in case traffic flows only
horizontally or vertically (so that even a continuous diagonal curve
is understood as a limit of many horizontal and vertical displacements).
In road traffic, this corresponds to a Manhattan-like network,
see \cite{Dafermos}. In the context of sensor networks this 
would correspond to directional antennas (either horizontal or vertical).
An alternative approach based on road traffic tools that is adapted to
omni-directional antennas can be found in \cite{road1,road2}
and we call it the $L2$-norm (meaning that the cost at any point
depends on the absolute value of the traffic there and not on its 
direction).
An extensive discussions on methods for numerical solutions of
our problem as well as the problem in \cite{road1} can be found
in \cite{road2} and in references there in, as well as in  \cite{Dafermos}.

\subsection{\bf The model}
Nodes are placed (deterministically) in a large number.
For energy efficiency, it is assumed that each node is equipped
with one or with two  directional antennas, allowing transmission
at each hop to be directed either from North to South or from West to East.
The model we use
extends that of \cite{Dafermos} to the multi-class framework.
We thus consider $\nu$ classes of
flows $ T_1^j \geq 0,\ T_2^j \geq 0 $, $j=1,...,\nu$.
To be compatible with Dafermos \cite{Dafermos}, we use her
definitions of orientation according to which the directions
North to South and West to East are taken positive.
In the dense limit, a curved path can be viewed as a limit of a path
with many such hops as the hop distance tends to zero.

Some assumptions on the cost:
\begin{itemize}
\item
{\bf Individual cost:}
We allow the cost for a horizontal transmission
(West-to-East, or equivalently, in the direction of the axis~$x_1$)
to be different than the cost for a vertical transmission
(North-to-South, or equivalently, in the direction of the axis~$x_2$).
It is assumed that a packet traveling in the direction of the axis $x_1$ incurs 
a {\sl transportation cost} $g_1$, and equivalently, traveling in the direction of the axis $x_2$ incurs 
a {\sl transportation cost} $g_2$.
Notice that the transportation costs $g_1$ and $g_2$ depend on the location $\mathbf x$
and the traffic flow $\mathbf T(\mathbf x)$ that is flowing through that location, {\it i.e.}, $g_1=g_1(\mathbf x,\mathbf T(\mathbf x))$
and $g_2=g_2(\mathbf x,\mathbf T(\mathbf x))$.

\item
We consider a {\sl vector transportation cost}
$ \mathbf g:=(g_1,g_2)$.
Notice that as each of its components, such vector transportation cost depends also upon the location $\mathbf x$ and the traffic flow $\mathbf T(\mathbf x)$
flowing through that location,
{\it i.e.}, $\mathbf g=\mathbf g(\mathbf x,\mathbf T(\mathbf x))$.

\item
The {\sl local transportation cost}~$g$ for the global optimization
problem is given by the inner product between the vector transportation cost and the flow of information, {\it i.e.},
\[
g(\mathbf x,\mathbf T(\mathbf x))=
\mathbf g(\mathbf x,\mathbf T(\mathbf x)) \cdot
\mathbf T(\mathbf x)=g_1 T_1+g_2 T_2,
\]
and it corresponds to the sum of the transportation costs multiplied by the quantity of flow in each direction.
\item
The {\sl global transportation cost} is the integral of the local transportation cost over the domain, {\it i.e.},
$\int_{\Omega}g(\mathbf x,\mathbf T(\mathbf x))\,d\x$.
\item
The local cost $ g (\mathbf x,\mathbf T(\mathbf x)) $
is assumed to be non-negative, 
monotone increasing in each component of $\mathbf T$
($T_1$ and $T_2$ in our $2$-dimensional case).
\end{itemize}

The {\bf boundary conditions} will be determined by the options
that travelers have in selecting their origin and/or destinations.
Examples of the boundary conditions are:
\begin{itemize}
\item {\sl Assignment problem}: users of the network have predetermined origin and destinations and are free to choose their travel paths.
\item {\sl Combined distribution and assignment problem}: users of the network have predetermined origins and are free to choose
their destinations (within a certain destination region) as well as their paths.
\item {\sl Combined generation, distributions and assignment problem}: users are free to choose their origins, their destinations,
as well as their travel paths.
\end{itemize}
The problem formulation is again to minimize $Z$ as
defined in (\ref{genOpt}). The natural choice of functional spaces to make
that problem precise, and to take advantage of the available theory developped
in the PDE (Partial Differential Equations) community,
is to work in the Sobolev space $H^1$.
We define $L^2(\Omega)$ as the space of functions that are square-integrable, {\it i.e.},
$L^2(\Omega)=\{f\text{ such that }\int_{\Omega}\lvert f(\x)\rvert^2\,d\x<+\infty\}$.
We define $H^1(\Omega)$ as the space of functions that are square integrable,
and with weak gradient 
square-integrable.
Take  $f$ to be a scalar function, define its weak
\emph{gradient} as a vector function $g$ such that, for any smooth 
vector function $\phi$ with compact support in $D$, $\int_D f\ \nabla\cdot\phi =
- \int_D <g, \phi>$.
Then, we seek $T^j_i$ in $H^1(\Omega)$,
such that $\rho$ is in $L^2(\Omega)$.
\bigskip

\subsection{\bf Karush-Kuhn-Tucker conditions}
The Karush-Kuhn-Tucker conditions (also known as KKT conditions)
which we introduce below
are necessary conditions for a solution in nonlinear programming to be optimal.
It is a generalization of the method of Lagrange multipliers.
We recall that the method of Lagrange multipliers provides
a strategy for finding the minimum of a function subject to constraints
and it is based on introducing new variables called Lagrange multipliers
and study a function called Lagrange function.
If a traffic flow~$\mathbf T$ 
is optimal for our global optimization problem 
then there exist Lagrange multipliers that satisfy some
complementarity conditions such that the corresponding 
Lagrangian is maximized by $\mathbf T$.

In the problem considered here the traffic flow in each direction~$T_i$
is a functional (a map from the vector space to the scalar space).
Then we will consider variational inequalities,
{\it i.e.}, inequalities involving a functional
which have to be solved for all the values of 
the vector space.

A key application of these KKT conditions will 
be introduced in Section~\ref{user_opt}. We 
study there the individual (non-cooperative) optimization
problem for each packet sent through the network
and show that the solution must satisfy some variational
inequalities. We then show that these inequalities
can be interpreted as the Karush-Kuhn-Tucker conditions
that we introduce in this chapter, applied to
some {\it transformed} cost (called ``potential'' and introduced by
Beckmann~\cite{beck}).
This will allow us to propose a method for solving
the individual (non-cooperative) optimization problem.

We begin by recalling Green's Theorem which has proved to be useful
for many physical phenomena. We will make extensive use of this theorem in this section
as well as in Section~\ref{user_opt}.
\begin{theorem}[Green's Theorem]\label{greenthm}
Let $\Omega$ be a region of the space, and let $\partial\Omega$ be its piecewise-smooth boundary.
Consider the scalar function $u$ and a continuously differentiable vector function $\mathbf v$, then
\[
\int_D u \nabla\cdot\mathbf v dx = \int_{\partial D} u <\mathbf v,\mathbf n> d\ell - \int_D <\mathbf v,\nabla u> dx.
\]
\end{theorem}

By making use of this theorem and the Karush-Kuhn-Tucker conditions
we are able to prove a result that provides a characterization of
the optimal solution for some special cases as we will see in the following.

\begin{theorem}
Define the Lagrangian as
\begin{equation*}
L^\zeta (\mathbf T):=
\int_\Omega\ell^\zeta (\mathbf x ,\mathbf T )\,d\mathbf x
\quad\text{with}\quad
\ell^\zeta (\mathbf x ,\mathbf T ):=
g(\mathbf x ,\mathbf T) -
\sum_{j=1}^\nu
\zeta^j (\mathbf x ) \Big[ \mathbf{\nabla}\!\cdot\!\mathbf T^j (\mathbf x ) -
\rho^j (\mathbf x ) \Big]
\end{equation*}
where $\zeta^j ( \mathbf x )\in L^2(\Omega)$ are called Lagrange
multipliers.

For a vector field $\mathbf T(\cdot)$
with positive components satisfying (\ref{conservation}), a necessary and sufficient condition for 
minimizing the cost (\ref{genOpt}) is that the Lagrangian be minimized over all
vector fields with positive components, or equivalently, that equations
\begin{subequations}
\begin{align}
\frac{ \partial g (\mathbf x ,\mathbf T ) }{ \partial T_i^j }
 + \frac{ \partial \zeta^j (\mathbf x ) }{ \partial x_i } &= 0 \quad
\mbox{ if }\quad T_i^j (\mathbf x ) > 0,\label{kkt1}\\
\frac{ \partial g (\mathbf x ,\mathbf T ) }{ \partial T_i^j }
 + \frac{ \partial \zeta^j (\mathbf x ) }{ \partial x_i } &\geq 0 \quad
\mbox{ if }\quad T_i^j (\mathbf x ) = 0.\label{kkt2}
\end{align}
\end{subequations}
be satisfied. 

\end{theorem}

{\bf Proof.-}
The criterion is convex, and the constraint
(\ref{conservation}) affine. Therefore the Karush-Kuhn-Tucker theorem
holds, stating that the Lagrangian is minimum at the optimum.
%
A variation $\delta\T(\cdot)$ will be admissible if
$\T(\x)+\delta\T(\x) \ge 0$ for all $\x$, hence in particular,
for all $\x$ such that $T_i^j(\x) = 0$ and $\delta T_i^j(\x)\ge 0$.

As we are working with functionals, we need a generalisation of the concept of directional derivative used in differential calculus.
The G\^ateaux differential $DF(u,d)$ of functional $F$ at $u$ in the direction $d$ is defined as
\[
DF(u,d)=\lim_{t\to 0}\frac{F(u+td)-F(u)}{t}=\frac{d}{dt}F(u+td)\Big|_{t=0}.
\]
if the limit exists. If the limit exists for all~$d$, one says that~$F$ is G\^ateaux differentiable at~$u$.

Let $\mathrm{D}L^\zeta$ denote the G\^ateaux derivative of functional
$L^\zeta$ with respect to $\T(\cdot)$. 
First order condition for local minimum reads
\[
\text{For all }\delta\T \>\mbox{admissible}\>, \mathrm{D}L^\zeta\cdot\delta\T \ge 0\,,
\]
therefore here
\[
\int_\Omega\sum_j\scal{\nabla_{\T^j}g(\x,\T(\x)),\delta\T^j(\x)}\,d\x -
\int_\Omega\sum_j\zeta^j(\x)\nabla\!\cdot\!\delta\T^j(\x)\,d\x \ge 0.
\]
Integrating by parts using Green's Theorem, this is equivalent to
\[
\int_\Omega\sum_j\left[\scal{\nabla_{\T^j}g,\delta\T^j}+
\scal{\nabla_\x\zeta^j,\delta\T^j}\right]d\x
-\int_{\partial\Omega}\sum_j\zeta^j\scal{\delta\T^j,\mathbf{n}}\,d\ell \ge 0\,.
\]
We may choose all the components $\delta\T^k = 0$ except $\delta\T^j$, and choose
$\delta\T^j$ in $(H^1_0(\Omega))^2$, {\it i.e.}, functions in $H^1(\Omega)$ such that their boundary integral be zero.
This is always feasible and admissible. 
Then the last term above vanishes,
and it is a classical fact that the inequality implies 
(\ref{kkt1})-(\ref{kkt2}) for $i=1,2$.

Placing this back in Euler's inequality,
and using a $\delta\T^j$ non zero on the boundary, it follows 
that necessarily\footnote{This is a complementary slackness condition 
on the boundary.}
$\zeta^j(\x) = 0$ at any $\x$ of the boundary $\partial\Omega$ where
$T(\x) > 0$. As we shall see this conditions provides the boundary
condition to recover the Lagrange multipliers $\zeta^j$ from equation (\ref{conservation}).

Equation (\ref{kkt1})-(\ref{kkt2}) is already stated in \cite{Dafermos} for
the single class case. However, as Dafermos states explicitly, its
rigorous derivation is not available there.
\endpf

Consider the following special cases that we shall need later. We
assume a single traffic class, but this could easily be extended to
several. Let
\begin{equation*}
g(\mathbf x,\mathbf T(\mathbf x))
= \sum_{i=1,2} g_i ( \mathbf x , \mathbf T ( \mathbf x ) )
 T_i ( \mathbf x ).
\end{equation*}
\begin{enumerate}
\item
Monomial cost per packet:
\begin{equation}\label{monom}
g_i (\mathbf x, \mathbf T(\mathbf x))=
k_i ( \mathbf x ) \Big( T_i ( \mathbf x ) \Big)^\beta
\end{equation}
for some $\beta>1$.
Then (\ref{kkt1})-(\ref{kkt2}) simplify to
\begin{subequations}
\begin{align}
(\beta+1)
k_i ( \mathbf x )
\left(
T_i (\mathbf x) \right)^\beta
 + \frac{ \partial \zeta (\mathbf x ) }{ \partial x_i } &= 0 \quad
\mbox{ if } T_i (\mathbf x ) > 0,\label{kkt5}\\
(\beta+1)
k_i (\mathbf x)
\left(
T_i(\mathbf x)\right)^\beta
 + \frac{ \partial \zeta (\mathbf x ) }{ \partial x_i } &\geq 0 \quad
\mbox{ if } T_i (\mathbf x ) = 0.\label{kkt6}
\end{align}
\end{subequations}
In that case, recovery of $\zeta$ to complete the process is difficult,
at best. Things are simpler in the next case.

\item
Affine cost per packet:
\begin{equation}
g_i ( \mathbf x , \mathbf T ( \mathbf x ) )  =
\frac{1}{2}k_i ( \mathbf x )  T_i ( \mathbf x )  + h_i ( \mathbf x )  .
\label{linearCq}
\end{equation}
Then (\ref{kkt1})-(\ref{kkt2}) simplify to
\begin{subequations}
\begin{align*}
k_i ( \mathbf x ) T_i(\x) + h_i ( \mathbf x )
 + \frac{ \partial \zeta (\mathbf x ) }{ \partial x_i } &= 0 \quad
\mbox{ if } T_i (\mathbf x ) > 0,\\
k_i ( \mathbf x ) T_i(\x) + h_i ( \mathbf x )
 + \frac{ \partial \zeta (\mathbf x ) }{ \partial x_i } &\geq 0 \quad
\mbox{ if } T_i (\mathbf x)=0.
\end{align*}
\end{subequations}
Assume that the $k_i(\cdot)$ are everywhere positive and bounded away from 0.
For simplicity, let $a_i=1/k_i$, and $b$ be the vector with coordinates
$b_i = h_i/k_i$, all assumed to be square integrable. Assume that there
exists a solution where $T(\x) > 0$ for all $\x$. Then
\[
T_i(\x)=-\left(a_i(\x)\frac{\partial\zeta(\x)}{\partial x_i}+b_i(\x)\right).
\]
As a consequence, from (\ref{conservation}) and the above remark, we get
that $\zeta(\cdot)$ is to be found as the solution in $H^1_0(\Omega)$ of the
elliptic equation (an equality in $H^{-1}(\Omega)$)
\[
\sum_i\frac{\partial}{\partial x_i}
\left(a_i(\x)\frac{\partial\zeta(\x)}{\partial x_i}\right) +\nabla\!\cdot\! b(\x)
+\rho(\x) = 0\,.
\]
This is a well behaved Dirichlet problem, known
to have a unique solution in $H^1_0(\Omega)$, furthermore easy to compute
numerically.
\end{enumerate}

\section{User optimization and congestion independent costs}

In this section, we extend the shortest path approach for optimization
that has already appeared using geometrical optics tools
\cite{geometry}.
We present a general optimization framework for handling shortest
path problems and more generally, minimum cost paths.

We consider the model of Section \ref{direct}. We assume
that the local cost depends on the direction of the flow but
not on its size. The cost is $c_1(\mathbf x)$ for a flow
that is locally horizontal and is
$c_2( \mathbf x)$ for a flow that is locally vertical.
We assume in this section that
$c_1$ and $c_2$ do not depend on $\mathbf T$.
The cost incurred by a packet
transmitted along a path $p$ is given by the line integral
\begin{equation}
\label{costeq}
\mathbf{c}_p=\int_p\mathbf{c}\cdot\,d\mathbf{x}.
\end{equation}

Let $V^j(\mathbf x)$ be the minimum cost to go from a point $\mathbf x$
to a set $B^j$, $j=1,...,\nu$. Then
\begin{equation}\label{dp1b}
V^j(\mathbf x)=\min\left(
c_1(\mathbf x)\,d x_1 + V^j ( x_1 + \,d x_1 , x_2),
c_2(\mathbf x)\, d x_2 + V^j ( x_1 , x_2 + \,d x_2)
\right).
\end{equation}
This can be written as the Hamilton Jacobi Bellman (HJB) equation:
\begin{equation}\label{dp1a}
0=\min\left(
c_1(\mathbf x)+\frac{\partial V^j(\mathbf x)}{\partial x_1},
c_2(\mathbf x)+\frac{\partial V^j(\mathbf x)}{\partial x_2}
\right), \qquad \forall\x \in B^j\,,\> V^j(\x) = 0\,.
\end{equation}
If $V^j$ is differentiable then, under suitable conditions,
it is the unique solution of (\ref{dp1a}).
In the case that $V^j$ is not everywhere differentiable
then, under suitable conditions, it is the unique
viscosity solution of (\ref{dp1a}) (see \cite{bardi,soner}).

There are many numerical approaches for solving the Hamilton-Jacobi-Bellman
(HJB) equation.
One can discretize the HJB equation and obtain a discrete dynamic programming
for which efficient solution methods exist. If one repeats this for
various discretization steps, then we know that the solution of
the discrete problem converges to the viscosity solution of the original
problem (under suitable conditions) as the step size
converges to zero \cite{bardi}.

\renewcommand{\xi}{\mathcal P}

\subsection{Geometry of minimum cost paths}



\begin{figure}[tbh]
\centering
\begin{minipage}[t]{70mm}
\includegraphics[height=2.0in,width=2.2in ]{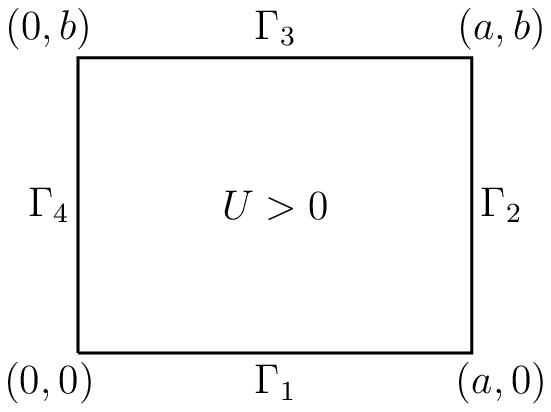}
 \caption{The rectangle~$R$ defined by the boundaries~$\Gamma_1\cup\Gamma_2\cup\Gamma_3\cup\Gamma_4$. The case where $U>0 $.}
\label{fig2G}
\end{minipage}
\hspace{0.5cm}
\begin{minipage}[t]{70mm}
\includegraphics[height=2.0in,width=2.2in ]{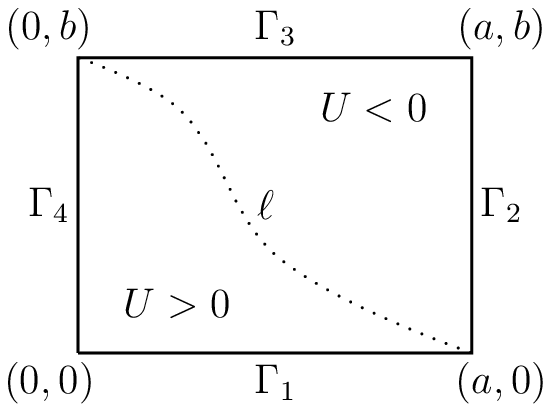}
 \caption{The case of two regions separated by a curve. Case \ref{np}.}
\label{fig3G}
\end{minipage}
\end{figure}

We consider now our directional antenna model in a
given rectangular area~$R$, 
defined by the simple closed
curve 
$\Gamma_1\cup\Gamma_2\cup\Gamma_3\cup\Gamma_4$
(see Fig.~\ref{fig2G}).
We study the case where transmissions can go from North to South
or from West to East.

We obtain below {\bf optimal paths} defined as paths
that achieve the minimum packet transmission cost defined by (\ref{costeq}).
We shall study two problems:
\begin{itemize}
\item
{\bf Point to point optimal path:}
we seek the minimum cost path between two points.
\item
{\bf Point to boundary optimal path:}
we seek the minimum cost path on a given region
that starts at a given point and is allowed to end at any point
on the boundaries.
\end{itemize}

Another formulation of Green's Theorem stated previously as Theorem~\ref{greenthm} 
give us a characterization of the optimal paths
for those two problems.

\begin{theorem}[Green's Theorem: alternative version]
Let $\Omega$ be a region of the space, and let $\partial\Omega$ be its piecewise-smooth boundary.
Suppose that $P$ and $Q$ are continuously differentiable functions in~$\Omega$.
Then
\[
\oint_{\partial\Omega} P dx+Q dy
=\int_{\Omega} \left( \frac{\partial Q}{\partial x}
-\frac{\partial P}{\partial y} \right) dx dy.
\]
\end{theorem}

Recall that the cost is composed of a horizontal and a vertical
component (these are $c_1( {\bf x } )$ and $c_2( {\bf x} ) $ 
respectively), which are constant (do not depend on the flow size).

Consider the function
\begin{equation*}
U(\mathbf x)=\frac{\partial c_2}{\partial x_1}
(\mathbf x)-\frac{\partial c_1}{\partial x_2}
(\mathbf x).
\end{equation*}

It will turn out that the structure of the minimum cost path
depends on the costs through the sign of the function $U$. Now,
if the function $\mathbf c$ is continuously differentiable then $U$ is a
continuous function. This motivates us to study cases in which
$U$ has the same sign everywhere (see Fig.~\ref{fig2G}),
or in which there are two regions in the rectangle $R$, one with $U>0$
and one with $U<0$, separated by a curve~$\ell$ on which~$U=0$
({\it e.g.} Fig.~\ref{fig3G}).

We shall assume throughout
that the function $\mathbf c$ is continuously differentiable, and that, if non-empty,
the set of points inside the domain where the function~$U$ is zero, {\it i.e.}, 
\mbox{$\ell = \{\x\,:\, U(\x)=0\}$}, is a smooth line. (This is true, {\it e.g.}, if
$\mathbf{c}$ is a smooth function and $\nabla U \neq 0$ on $\ell$.)

\subsection{The function $U$ has the same sign over the 
whole region}\label{samesign}

\begin{figure}[tbh]
\begin{center}
\subfigure[Case $U>0$.]{
\includegraphics[height=2.5in,width=2.8in]{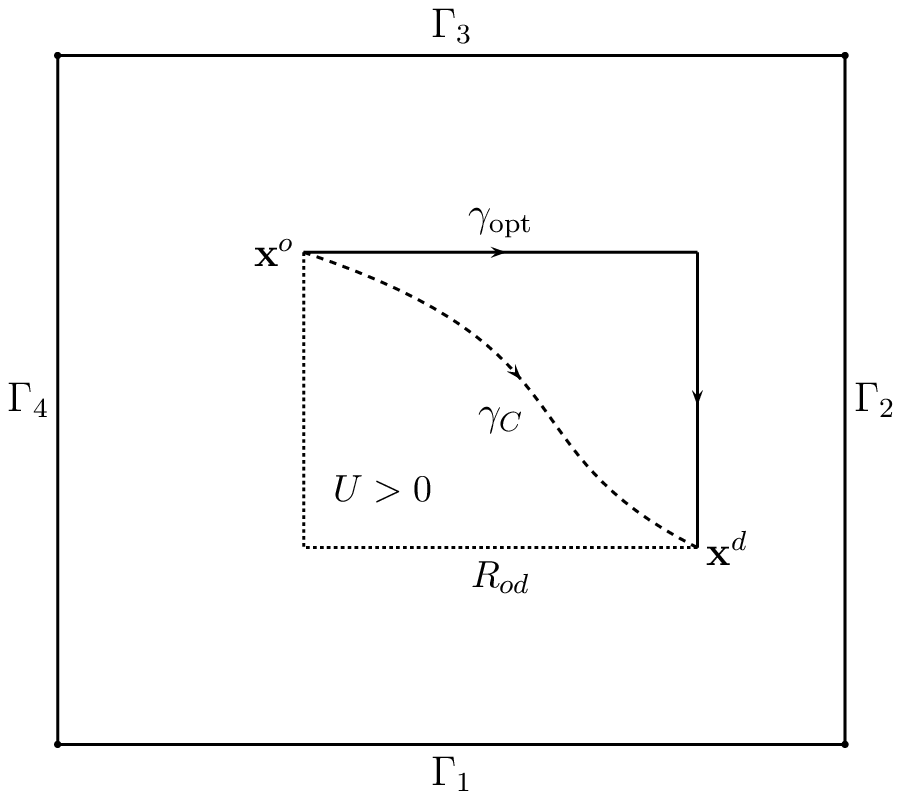}
\label{fig1P}
}
\hspace{0.05in}
\subfigure[Case $U<0$.]{
\includegraphics[height=2.5in,width=2.8in]{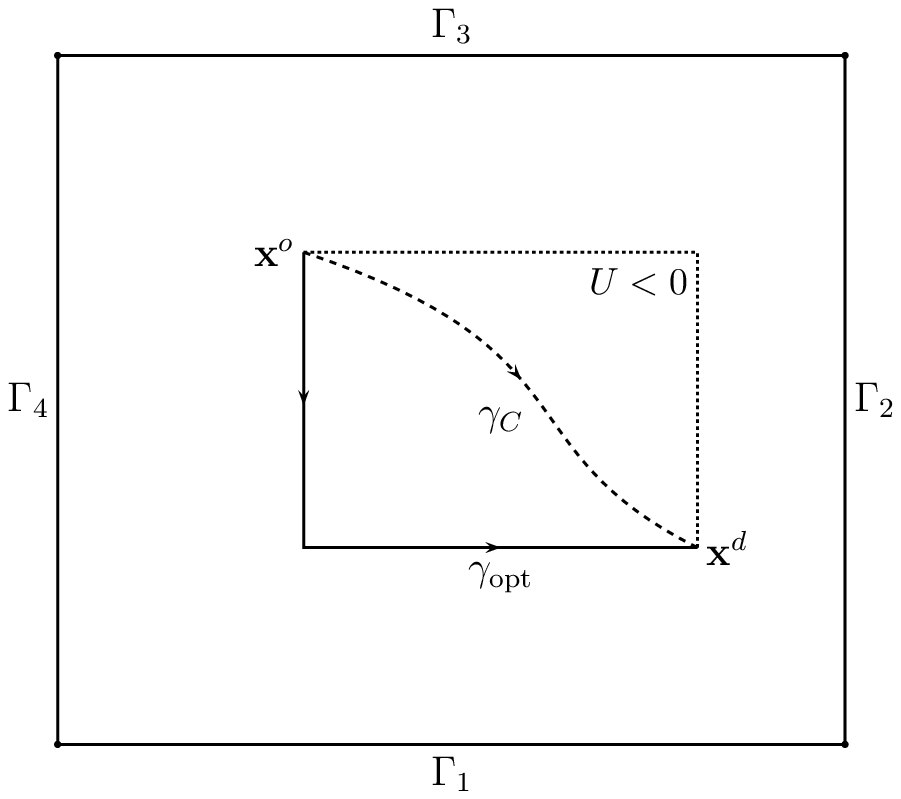}
\label{fig2P}
}
\hspace{0.2in}
\caption{Optimal paths when $U>0$ \ref{fig1P} and when $U<0$ \ref{fig2P} in the interior rectangle defined by the origin point and the destination point.}
\end{center}
\end{figure}

\begin{theorem}\label{theorem:positive}
(Point to point optimal path)
Suppose that an origin point
$\mathbf x^o=(x_1^o,x_2^o)$
wants to send a packet
to a destination point $\mathbf x^d=(x_1^d,x_2^d)$
and both points are in the interior of rectangle~$R$.
\begin{enumerate}[i.]
\item\label{UpositivoOD}
If the function $U$ is positive almost everywhere in the interior rectangle~$R_{od}$ defined by both points (see Fig.~\ref{fig1P}), 
then the optimal path~$\gamma_{\rm opt}$ is given by a horizontal straight line~$\gamma_H$ and then a vertical straight line~$\gamma_V$ (see Fig.~\ref{fig1P}).

More precisely,~$\gamma_{\rm opt}=\gamma_H\cup\gamma_V$ where
\begin{gather*}
\gamma_H=\{(x_1,x_2)\text{ such that }x_1^o\leq x_1\leq x_1^d, x_2=x_2^o\},\\
\gamma_V=\{(x_1,x_2)\text{ such that }x_1=x_1^d, x_2^o\leq x_2\leq x_2^d\}.
\end{gather*}
\item\label{UnegativoOD}
If the function $U$ is negative almost everywhere in the interior rectangle~$R_{od}$
then there is an optimal path~$\gamma_{\rm opt}$ given by a vertical straight line~$\gamma_V$ and then
a horizontal straight line~$\gamma_H$ (see Fig.~\ref{fig2P}).

More precisely,~$\gamma_{\rm opt}=\gamma_V\cup\gamma_H$ where
\begin{gather*}
\gamma_V=\{(x_1,x_2)\text{ such that } x_1=x_1^o, x_2^o\leq x_2\leq x_2^d\},\\
\gamma_H=\{(x_1,x_2)\text{ such that } x_1^o\leq x_1\leq x_1^d, x_2=x_2^d\}.
\end{gather*}
\item\label{CaminoUnico}
In both cases, $\gamma^{\text{opt}}$ is unique almost surely
({\it i.e.}, the area between $\gamma^{\text{opt}}$
and any other optimal path is zero).
\end{enumerate}
\end{theorem}

{\bf Proof.-}
Consider an arbitrary path\footnote{Respecting that
each subpath can be decomposed
in sums of paths either from North to South or from West to East
(or is a limit of such paths).
}
$\gamma_C$
joining $\mathbf x^o$ to $\mathbf x^d$, and assume that the Lebesgue
measure of the area between $\gamma^{\text{opt}}$
and $\gamma_C$ is nonzero.
We call~$\gamma_C$ the comparison path (see Fig.~\ref{fig1P}
for the case $U>0$ and Fig.~\ref{fig2P} for the case $U<0$).


(\ref{UpositivoOD}) Showing that the cost over path $\gamma_{\rm opt}$ is optimal is
equivalent to showing that the integral of the cost over the closed path
$\xi$ is negative. Hereby $\xi$ is
given by following $\gamma^{\text{opt}}$ from the origin $\mathbf x^o$
to the destination $\mathbf x^d$, and then returning from the destination $\mathbf x^d$
to the origin $\mathbf x^o$
by moving along the comparison path $\gamma_C$ in the reverse direction.
This closed path is written as
$\xi=\gamma_{\rm opt}\cup\gamma_C^-$ and $A$ denotes the bounded area
described by $\xi$. Using Green's Theorem we obtain
\begin{equation*}
\oint_{\xi}\mathbf c\cdot\mathbf{dx}=-\int_A U(\mathbf x)dS
\end{equation*}
which is strictly negative since $U>0$ almost everywhere~on the interior rectangle~$R_{od}$.
Decomposing the left integral, this concludes the proof of (i), and establishes at the same time
the corresponding statement on uniqueness in (\ref{CaminoUnico}).
\\
(\ref{UnegativoOD}) is obtained similarly.
\endpf

\begin{figure}[tbh]
\centering
\begin{minipage}[t]{70mm}
\includegraphics[height=2.0in,width=2.2in]{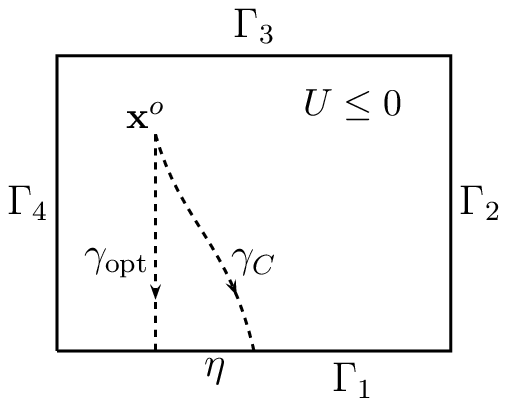}
\caption{Theorem \ref{lemma:negative} (\ref{Unegativo})}
\label{f62i}
\end{minipage}
\hfill
\begin{minipage}[t]{70mm}
\includegraphics[height=2.0in,width=2.2in]{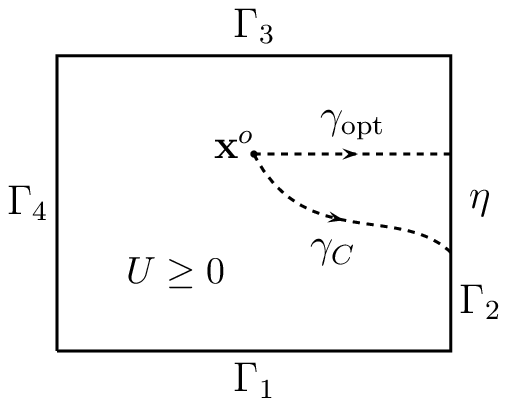}
\caption{Theorem \ref{lemma:negative} (\ref{Upositivo})}
\label{f62ii}
\end{minipage}
\end{figure}

\begin{theorem}\label{lemma:negative}
(Point to boundary optimal path)\\
Consider the problem of finding an optimal path from
a point $\mathbf x^o$ in the rectangle~$R$ to the boundary
$\Gamma_1\cup\Gamma_2$.

\begin{enumerate}[i.]
\item\label{Unegativo} If the function~$U$ is almost everywhere negative inside the rectangle~$R$ 
and the cost on the boundary $\Gamma_1$ is non-negative and
on boundary $\Gamma_2$ is non-positive, then the optimal path
is the straight vertical line (See Fig.~\ref{f62i}).
\item\label{Upositivo} If the function~$U(\mathbf x)$ is almost everywhere positive inside the rectangle~$R$ 
and the cost on the boundary $\Gamma_1$ is non-positive and
on boundary $\Gamma_2$ is non-negative. Then the optimal path
is the straight horizontal line (See Fig.~\ref{f62ii}).
\end{enumerate}
\end{theorem}

{\bf Proof.-}

(\ref{Unegativo}) Denote by $\gamma_{\rm opt}$ the straight vertical path joining $\mathbf x^o$ to the boundary~$\Gamma_1$.
Consider another arbitrary valid path $\gamma_C$ joining $\mathbf x^o$
to any point ${\bf x}^*$ on the boundary
$\Gamma_1\cup\Gamma_2$, and assume that the Lebesgue measure
of the area between $\gamma_{\text{opt}}$ and $\gamma_C$ is nonzero.
We call~$\gamma_C$ the comparison path.

Assume first that
${\bf x}^*$ is on the boundary~$\Gamma_2$.  Denote
$ {\bf x}^D$ the South-East corner of the rectangle~$R$, {\it i.e.}, ${\bf x}^D:= \Gamma_1\cap\Gamma_2$.
Then by Theorem \ref{theorem:positive}(ii), the cost to go from
$\mathbf x^o$ to $\mathbf x^d $ is smaller when using $\gamma_{\rm opt}$
and then continuing eastwards (along $\Gamma_1$) than when using the comparison path~$\gamma_C$
and then southwards (along $\Gamma_2$). Due
to our assumptions on the costs over
the boundaries, this implies that the cost along the straight vertical path~$ \gamma_{\rm opt}$ is smaller
than along the comparison path~$\gamma_C$.

Next consider the case where $\mathbf x^*$ is on the boundary~$\Gamma_1$.
Denote by $\eta$ the section of the boundary  $\Gamma_1$
that joins $\gamma_{\rm opt} \cap \Gamma_1$ with $\mathbf x^*$  (see
Figure \ref{f62i}).
Then again, by Theorem \ref{theorem:positive} (ii), the cost to go from
$\mathbf x^o$ to $\mathbf x^* $ is smaller when using $\gamma_{\rm opt}$
and then continuing eastwards (along $\Gamma_1$) than when using
the comparison path~$\gamma_C$. Due to our assumptions that the cost on
$\Gamma_1$ is non-negative, this implies that the cost
along $ \gamma_V$ is smaller than along $\gamma_C$.

(\ref{Upositivo}) is obtained similarly.
\endpf

\subsection{The function $U$ changes sign within the region $R$}

Consider the region on the space
$\ell:=\left\{\mathbf x\in R\text{ such that }U(\mathbf x)=0\right\}.$
Let us consider the case when~$\ell$ is a valid path in the rectangular area, such that it starts
at the North-West corner (the intersection of the boundaries $\Gamma_3\cap\Gamma_4$) and finishes at
the South-East corner (the intersection of the boundaries $\Gamma_1\cap\Gamma_2$).
Then the space is divided in two areas, and as the function $U$ is
continuous we have the following cases:

\begin{enumerate}
\item\label{np} $U(\mathbf x)$ is negative in the upper area
and positive in the lower area
(see Fig.~\ref{fig3G}).
\item\label{pn} $U(\mathbf x)$ is positive in the
upper area and negative in the lower area.
\end{enumerate}
The two other cases where the sign of~$U$ is the same over~$R$ are
contained in what we solved in the previous subsection~\ref{samesign}.
\bigskip

{\bf Case \ref{np}:} The function $U(\mathbf x)$ is
negative in the upper area and positive in the lower area.

We shall show that in this case, $\ell$~is an attractor, in the sense that the optimal path reaches the line~$\ell$
with the minimal possible distance and then continues along this line until it reaches the destination.

\begin{proposition}\label{A1}
Assume that the origin point $\mathbf x^o$ and the destination point $\mathbf x^d$ are both on~$\ell$.
Then the path $p_\ell$ that follows~$\ell$ from the origin point $\mathbf x^o$ to the destination point $\mathbf x^d$ is optimal.
\end{proposition}

{\bf Proof.-}
Consider a comparison path $\gamma_C$
that coincides with $\ell$ only in the origin~$\mathbf x^o$ and
destination~$\mathbf x^d$ points. First assume that the comparison path~$\gamma_C$ is entirely
in the upper ({\it i.e.}, northern) part and call $A$
the area between~$\gamma_C$ and $p_\ell$. Define $\xi$ to be the closed path
that follows $p_\ell$ from $\mathbf x^o$ to $\mathbf x^d$
and then returns along $\gamma_C$.

The integral
$ \int_A U(\mathbf x)\,d\x$ is negative by
assumption. By Green's Theorem, it is equal to~
\mbox{$ \oint_{\xi}\mathbf c\cdot\,d\x $}.
This implies that the cost along $p_\ell$ is strictly
smaller than along $\gamma_C$.

A similar argument holds for the case that  $\gamma_C$
is below $p_\ell$.

A path between $\mathbf x^o$ and $\mathbf x^d$ may have several intersections with $\ell$.
Between each pair of consecutive intersections of $\ell$, the subpath has a
cost larger than that obtained by following $\ell$ between these points (this
follows from the previous steps of the proof). We conclude that $p_\ell$
is indeed optimal.
\endpf

\begin{proposition}\label{A}
Let an origin point $\mathbf x^o $ send packets to a destination point ${\bf x}^D $.
\begin{enumerate}[i.]
\item\label{prop:Ai}
Assume both points are in the upper region.
Denote by $\gamma_1$
the two segments path given by Theorem \ref{theorem:positive} (ii).
Then the optimal curve $\gamma_{\rm opt}$ is 
obtained as the maximum between $\ell$ and $\gamma_1$\footnote{By
the maximum we mean the following:
If $\gamma_1$ does not intersect $\ell$, then $\gamma_{\rm opt}=\gamma_1$.
If it intersects $\ell$, then $\gamma_{\rm opt}$ agrees with $\gamma_1$
over the path segments where $\gamma_1$ is in the upper region
and otherwize agrees with~$\ell$. The minimum is defined similarly.}.
\item\label{prop:Aii}
Let both points be in the lower region.
Denote by $\gamma_2$
the two segments path given in Theorem \ref{theorem:positive} (i).
Then the optimal curve $\gamma_{\rm opt}$ is
obtained as the minimum between $\ell$ and~$\gamma_2$.
\end{enumerate}
\end{proposition}

{\bf Proof.-}

(\ref{prop:Ai}) A straightforward adaptation of the proof of the previous proposition
implies that the path in the statement of the proposition is optimal among
all those restricted to the upper region. Consider now a path $\gamma_C$
that is not restricted to the upper region. Then $\ell\cap\gamma_C$
contains two distinct points such that $\gamma_C$
is strictly lower than $\ell$ between these points. Applying
Proposition~\ref{A1}, we then see that the cost
of $\gamma_C$ can be strictly improved by following $\ell$ between
these points instead of following $\gamma_C$ there.
This concludes (\ref{prop:Ai}).

(\ref{prop:Aii}) Proved similarly.
\endpf

\begin{proposition}\label{A2}
Let a point $\mathbf x^o$ send
packets to a point $\mathbf x^d$.
\begin{enumerate}[i.]
\item\label{ul}
Assume the origin is in the upper region and the destination in
the lower one. Then the optimal path has three segments;
\begin{enumerate}[1.]
\item
It goes straight vertically from $\mathbf x^o $ to $\ell$,
\item
Continues as long as possible along $\ell$, {\it i.e.}, until it reaches the
first coordinate of the destination,
\item
At that point it goes straight vertically from $\ell$ to  $\mathbf x^d$.
\end{enumerate}
\item\label{lu}
Assume the origin is in the lower region and the destination in
the upper one. Then the optimal path has three segments;
\begin{enumerate}[1.]
\item
It goes straight horizontally from $\mathbf x^o $ to $\ell$,
\item
Continues as long as possible along $\ell$, {\it i.e.}, until it reaches the
second coordinate of the destination,
\item
At that point it goes straight horizontally from $\ell$ to  $\mathbf x^d $.
\end{enumerate}
\end{enumerate}
\end{proposition}

{\bf Proof.-}
The proofs of (\ref{ul}) and of (\ref{lu}) are the same.
Consider an alternative route $\gamma_C$. Let $ \tilde {\bf x} $
be some point in $\gamma_C \cap \ell$. The proof now follows by
applying the  previous proposition to obtain first the optimal path between
the origin and $\tilde {\bf x}$ and second, the optimal path between
$\tilde {\bf x}$ and the destination.
\endpf

\bigskip

{\bf Case \ref{pn}:} The function $U$ is positive in the upper area and negative in the lower area.

\bigskip


This case turns out to be more complex than the previous one.
The curve $M$ has some obvious repelling properties which we state
next, but they are not as general as the attractor properties that
we had in the previous case.

\begin{proposition}\label{C}
Assume that both origin and destination are in the same region.
Then the paths that were optimal in Theorem \ref{theorem:positive}
are optimal here as well, if we restrict ourselves to paths that remain
in the same region.
\end{proposition}

{\bf Proof.-} Given that the origin and destination are in a region
we may change the cost over the other region so that it has the same
sign over all the region $R$. This does not influence the cost of
path restricted to the region of the origin-destination pair. With
this transformation we are in the scenario of Theorem
\ref{theorem:positive} which we can then apply.
\endpf

{\bf Discussion}.-
Note that the (sub)optimal policies obtained in Proposition \ref{C}
indeed look like being repelled from $\ell$; their two segments trajectory
guarantees to go from the origin to the destination as far as possible
from $\ell$.

We note that
unlike the attracting structure that we obtained in Case 1,
one cannot extend the repelling structure to the case where the paths are
allowed to traverse from one region to another.

\bigskip

\section{User optimization and congestion dependent cost}
\label{user_opt}

We now go beyond the approach of the previous section by allowing
the cost to depend on congestion. Shortest path costs can
be a system objective as we shall motivate below. But it can
also be the result of decentralized decision making by many
``infinitesimally small" players where a player may represent a single packet
(or a single session) in a context where there is a huge population
of packets (or of sessions). The result of such a decentralized
decision making can be expected to satisfy the following properties
which define the so called, user (or Wardrop) equilibrium:

``{\em Under equilibrium conditions traffic arranges itself in congested
networks such that all used routes between an OD pair (origin-destination pair)
have equal and minimum
costs while all unused routes have greater or equal costs}'' \cite{Wardrop}.

{\bf Motivation}.-
One popular objective in some routing protocols in ad hoc
networks is to assign routes for packets in a way that
each packet follows a minimal cost path (given the others' paths
choices) \cite{GKcdc}.
This has the advantage of equalizing origin-destination
delays of  packets that belong to the same class, which allows one to
minimize the amount of packets that come out of sequence (this
is desirable  since in data transfers, out of order packets are
misinterpreted to be lost which results not only in retransmissions
but also in drop of systems throughput).

{\bf Related work}.-
Both the framework of global optimization as well as the one of
minimum cost path have been studied extensively
in the context of road traffic engineering. The use of a continuum
network approach was already introduced on 1952 by Wardrop \cite{Wardrop}
and by Beckmann \cite{Beckmann}.
For more recent papers in this area, see
{\it e.g.}~\cite{Dafermos,daniele02,hoWong,idone04,wongsens} and references
therein. We formulate it below and obtain some of its properties.


\noindent
{\bf Congestion dependent cost}.-
We allow the individual transmission cost~$c_1$ for a horizontal transmission (in the direction of the axis~$x_1$)
to be different than the individual transmission cost~$c_2$ for a vertical transmission (in the direction of the axis~$x_2$).
We add to the individual transmission cost $c_1$ the dependence on the traffic flow $T_1$ (in the direction of the axis~$x_1$) and to
the individual transmission cost $c_2$ the dependence on the traffic flow $T_2$ (in the direction of the axis~$x_2$), as we did in Section~\ref{direct} to the
transportation cost for the global optimization problem.

Let $V^j( \mathbf x )$ be the minimum cost to go from a point $\mathbf x$
to $B^j$ at equilibrium. Equation~(\ref{dp1b}) still holds
but this time with $c_i$ that depends on $T_1^j$, $T_2^j$,
and on the total flows $T_1$, $T_2$.
Thus (\ref{dp1a}) becomes, for all $j \in\{1,\ldots,\nu\}$,
\begin{equation}\label{dp1c}
 0 = \min_{i=1,2} \Big(
c_i( \mathbf x, T_i )   + \frac{ \partial  V^j ( \mathbf x ) }{ \partial x_i }
\Big)\,,\quad \forall \x\in B^j\,,V^j(\x) = 0\,.
\end{equation}
Notice that this method can be viewed as a generalization of the optimization method known as dynamic programming,
in particular, last equation would be a generalization of Bellman equation also known as dynamic programming equation.

We note that if $T_i^j ( \mathbf x  ) > 0$ then by the definition of
the equilibrium, $i$ attains the minimum at (\ref{dp1c}).
Hence (\ref{dp1c}) implies the following relations for each traffic class $j$,
and for $i=1,2$:
\begin{subequations}
\begin{align}
\label{wrdrp1}
& c_i ( \mathbf x , T_i )
+\frac{\partial V^j}{\partial x_i}=0\quad\textrm{if}\quad T_i^j >0,\\
& c_i ( \mathbf x , T_i )  +
\frac{\partial V^j}{\partial x_i}\geq 0\quad\textrm{if}\quad T_i^j=0.
\label{wrdrp2}
\end{align}
\end{subequations}
This is a set of coupled PDE's (Partial Differential Equations), 
actually difficult to analyse further.

\noindent
{\bf Beckmann transformation} \\
As Beckmann {\it et al.} did in \cite{beck} for discrete networks, we
transform the minimum cost problem into an equivalent global
minimization one. We shall restrict our analysis to the single class
case.  To that end, we note that equations (\ref{wrdrp1})-(\ref{wrdrp2})
have exactly the same form
as the Karush-Kuhn-Tucker conditions (\ref{kkt1})-(\ref{kkt2}), except that
$c_i(\mathbf x, T_i)$ in the former are replaced by
$ { \partial g (\mathbf x,\mathbf T)}/{ \partial T_i(\mathbf x) }  $
in the latter.
We therefore introduce a \emph{potential function} $\psi$ defined by
\begin{equation*}
\psi (\mathbf x , \mathbf T ) =
\sum_{i=1,2}
\int_0^{T_i}\! c_i ( \mathbf x , s  ) {\rm d} s
\end{equation*}
so that for both $i=1,2$:
\begin{equation*}
c_i (\mathbf x, T_i ) =
\frac{\partial\psi(\mathbf x,\mathbf T)}{ \partial T_i }\,.
\end{equation*}
Then the user equilibrium flow is the one obtained from the global
optimization problem where we use $  \psi (\mathbf x , \mathbf T ) $
as local cost. We conclude the following.

\begin{theorem}
Let $x^*$ be a solution to the following global optimization
problem.
\begin{equation*}
\min_{T(\cdot)} \int_\Omega
\psi (\mathbf x , \mathbf T )\,d \mathbf x
\qquad \mbox{subject to }
\mathbf{\nabla}\cdot\mathbf T (\mathbf x) =\rho
(\mathbf x), \qquad \forall\mathbf x\in\Omega.
\end{equation*}
Then it is the Wardrop equilibrium.
\end{theorem}

\begin{remark}
In the special case where costs are given as a power of the flow
as defined in eq. (\ref{monom}), we observe that equations
(\ref{wrdrp1})-(\ref{wrdrp2}) coincide with equations
(\ref{kkt5})-(\ref{kkt6}) (up-to a multiplicative constant of the cost).
We conclude that for such costs, the
user equilibrium and the global  optimization solution coincide.
\end{remark}

\section{Example}

The following example is an adaptation of
the road traffic problem solved by Dafermos in~\cite{Dafermos}
to our ad hoc setting. We therefore use the notation of~\cite{Dafermos}
for the orientation, as we did in Section \ref{direct}.
Thus the direction from North to South
will be our positive $x_1$ axis, and from West to East will be the positive
$x_2$ axis.
The framework we study is the user optimization problem with congestion dependent cost.
For each point on the West and/or North boundary we consider the
point to boundary problem. We thus seek a Wardrop equilibrium where each
user can choose its destination among a given set. A flow configuration
is a Wardrop equilibrium if under this configuration,
each origin chooses a destination and a path to that destination
that minimize that user's cost among all its possible choices.

Consider the rectangular area $R$ on the bounded domain $\Omega$
defined by the simple closed curve
$\partial R^+=\Gamma_1^+\cup\Gamma_2^+\cup\Gamma_3^-\cup\Gamma_4^-$
where
\[
\begin{array}{ll}
\Gamma_1=\left\{0\leq x_1\leq a,\quad x_2=0\right\}, &
\Gamma_2=\left\{x_1=a,\quad 0\leq x_2\leq b\right\},\\
\Gamma_3=\left\{0\leq x_1\leq a,\quad x_2=b\right\}, &
\Gamma_4=\left\{x_1=0,\quad 0\leq x_2\leq b\right\}.
\end{array}
\]
Assume throughout
that $\rho=0$ for all $ {\mathbf x}$ in the interior of $\Omega$, and that
the costs of the routes are linear, {\it i.e.},
\begin{equation}\label{segunda}
c_1=k_1 T_1+h_1\quad\textrm{and}\quad c_2=k_2 T_2+h_2,
\end{equation}
with $k_1>0$, $k_2>0$, $h_1$, and $h_2$ constant over $\Omega$.
Linear costs can be viewed as a Taylor approximation
of an arbitrary cost in the light traffic regime.

We are precisely in the framework of Section~\ref{direct} and Section \ref{user_opt}
with affine costs per packet. As a matter of fact,
the potential function associated with these costs is
\[
\psi(\T) = \sum_{i=_1}^2\int_0^{T_i}(k_is+h_i)\,ds =
\sum_{i=1}^2 (\frac{1}{2}k_iT_i+h_i)T_i\,.
\]
Now, we want to handle a condensation of origins or destinations along the
boundary. While this is feasible with the framework of section
\ref{direct}, it is rather technical. We rather use a more
direct path below.

Notice that in the interior of $\Omega$, we have
\begin{equation*}
\frac{\partial T_1}{\partial x_1}+\frac{\partial T_2}{\partial x_2}=0.
\end{equation*}

Take any closed path $\gamma$ surrounding a
region $\omega$. Then by Green formula,
\[
\oint_\gamma
T_1 d \xi_2 - T_2 d \xi_1 =
\int_\omega
\frac{ \partial T_1 }{ \partial x_1  } +
\frac{ \partial T_2 }{ \partial x_2  } = 0
\]
Therefore we can define
\[
\phi( {\bf x} ) :=
\int_{{\bf x^o}}^{\bf x}
T_1  d \xi_2
- T_2  d \xi_1
\]
the integral will not depend on the path between
$ {{\bf x^o}} $ and $ {\bf x} $ and $\phi$ is thus well defined,
and we have
\begin{equation}
\label{primera}
\frac {\partial \phi ( { \bf x} ) }{ \partial x_2 } =
T_1 ( { \bf x} )
\qquad
\frac {\partial \phi ( { \bf x} ) }{ \partial x_1 } =
- T_2 ( { \bf x} ) \,.
\end{equation}
%
%
%
We now make the assumption that there is sufficient demand and that
the congestion cost is not too high so that at equilibrium
the traffic $T_1$ and $T_2$ are strictly positive over all $\Omega$~\cite{Dafermos}.
It turns out that all paths to the destination
are used. Thus, from Wardrop's principle, the cost $\int\!\mathbf{c}\,dx$
is equalized between any two paths. And therefore,
\begin{equation*}
\frac{\partial c_1}{\partial x_2}=\frac{\partial c_2}{\partial x_1}.
\end{equation*}
Using the equations in (\ref{segunda}) then
\begin{equation*}
 k_1\frac{\partial T_1}{\partial x_2}=k_2\frac{\partial T_2}{\partial x_1},
\end{equation*}
and from equations in (\ref{primera}) we have
\begin{equation*}
k_1\frac{\partial^2\phi}{\partial x_2^2}+k_2\frac{\partial^2\phi}{\partial x_1^2}=0.
\end{equation*}
Let $k_i = K_i^2$. Divide the above equation by $k_1k_2$. One obtains
\[
\frac{1}{K_1^2}\frac{\partial^2\phi}{\partial x_1^2} +
\frac{1}{K_2^2}\frac{\partial^2\phi}{\partial x_2^2} = 0.
\]
Following the classical way of analyzing the Laplace equation,
(see\cite{weinberger}) we attempt a separation of variables
according to
\[
\phi(x_1,x_2) = F_1(K_1x_1)F_2(K_2x_2)\,.
\]
We then get that
\[
\frac{F_1''(K_1x_1)}{F_1(K_1x_1)} = -\frac{F_2''(K_2x_2)}{F_2(K_2x_2)} =
s^2\,.
\]
In that formula, since the first term is independent on $x_2$ and
the second on $x_1$, then both must be constant. We call $s^2$ that
constant, but we do not know its sign. Therefore, $s$ may be
imaginary or real. All solutions of this system for a given $s$ are
of the form
\[
F_1(x) = A\cos(isx)+B\sin(isx)\,,\qquad F_2 = C\cos(sx)+D\sin(sx)\,.
\]
As a matter
of fact, $\phi$ may be the sum of an arbitrary number of such multiplicative
decompositions with different $s$. We therefore arrive at the general formula 
\[
\phi(x_1,x_2) =
\!\int\![A(s)\cos(isK_1x_1)+B(s)\sin(isK_1x_1)]
[C(s)\cos(sK_2x_2)+D(s)\sin(sK_2x_2)]\,ds.
\]
From this formula, we can write $T_1$ and $T_2$ as integrals also.
The flow $T$ at the boundaries should be orthogonal to the boundary, and have
the local origin density for inward modulus (it is outward at a sink).
It remains to expand these boundary conditions in Fourier integrals to
identify the functions $A$, $B$, $C$, and $D$, which is tedious but straightforward 
(it is advisable to represent the integrals of the boundary densities as
Fourier integrals, since then the boundary conditions themselves will be of
the form $s\!\int\! R(s)\, ds$, closely matching the formulas we obtain for
the $T_i$'s).

\section{Conclusions}

Routing in ad hoc networks has received much attention in the
massively dense limit. The main tools to describe the limits had
been Electrostatics and geometric optics. We exploited another
approach for the problem that has its roots in road traffic theory,
and presented both quantitative as well as qualitative results for
various optimization frameworks. The links to road traffic theory
allow us to benefit of the results of more than 
fifty years of research in that area that not only provide
mature theoretical tools but have also advanced in numerical solution
methods.

The continuum topology that we used in this paper is to be viewed as
an approximation for dense networks. However, we believe that 
it may arise in other applications as well. As an example, consider
a standard routing problem, in which instead of having a fixed rate of
packets arrivals, we only have a constraint on the total amount
of arrivals of packets, and then one has to choose both the instantaneous
routes as well as at what (time-dependent) rate to transmit the packets.
This type of problem (also much studied in road traffic context)
can be viewed as routing over a continuum topology where time replaces
space.

\section*{Acknowledgement}
We wish to thank Dr. Stavros Toumpis for helpful discussions.
The work has been partially supported by the Bionets European Contract.
The first author was partially supported by CONICYT (Chile) and INRIA (France).
The first and fourth author were partially supported by Alcatel-Lucent
within the Alcatel-Lucent Chair in Flexible Radio at Supelec.

{\small



\end{document}